\documentclass[]{spie}  

\usepackage{amsmath,amsfonts,amssymb}
\usepackage{graphicx}
\usepackage[colorlinks=true, allcolors=blue]{hyperref}
\usepackage{blindtext}
\usepackage{enumitem}

\title{Optimal input configuration of dynamic contrast enhanced MRI in convolutional neural networks for liver segmentation}

\author{Mari\"elle J.A. Jansen}
\author{Hugo J. Kuijf}
\author{Josien P.W. Pluim}
\affil{Center for Image Sciences, University Medical Center Utrecht, the Netherlands}

\authorinfo{Corresponding author: M.J.A.Jansen-35@umcutrecht.nl}

\begin{document} 
\maketitle

\begin{abstract}
Most MRI liver segmentation methods use a structural 3D scan as input, such as a T1 or T2 weighted scan. Segmentation performance may be improved by utilizing both structural and functional information, as contained in dynamic contrast enhanced (DCE) MR series. Dynamic information can be incorporated in a segmentation method based on convolutional neural networks in a number of ways. In this study, the optimal input configuration of DCE MR images for convolutional neural networks (CNNs) is studied.\par
The performance of three different input configurations for CNNs is studied for a liver segmentation task. The three configurations are I) one phase image of the DCE-MR series as input image; II) the separate phases of the DCE-MR as input images; and III) the separate phases of the DCE-MR as channels of one input image. The three input configurations are fed into a dilated fully convolutional network and into a small U-net. The CNNs were trained using 19 annotated DCE-MR series and tested on another 19 annotated DCE-MR series. The performance of the three input configurations for both networks is evaluated against manual annotations.\par
The results show that both neural networks perform better when the separate phases of the DCE-MR series are used as channels of an input image in comparison to one phase as input image or the separate phases as input images. No significant difference between the performances of the two network architectures was found for the separate phases as channels of an input image.

\end{abstract}

\keywords{: Dynamic contrast enhanced MRI, liver, segmentation, deep learning}

\section{INTRODUCTION}
\label{sec:intro}  
Automatic liver segmentation in abdominal MR images is an important step in automatic detection and characterization of focal liver lesions. Most (semi-)automatic MR liver segmentation methods are designed to process an anatomical 3D image as input instead of the dynamic contrast enhanced (DCE) MR images.\cite{Lopez-Mir2014,Huynh2017,Chartrand2017,Masoumi2012} These DCE-MR images hold valuable information about liver shape and function in the different contrast phases. Using the rich information of the DCE-MR images in convolutional neural networks (CNNs) might improve the segmentation performance.\par
However, it is unclear how dynamic MR series are best incorporated in CNNs. There are several ways to process the series in a CNN. The conventional way is to use one of the 3D phase images of the DCE-MR series as input of a CNN, but this neglects the dynamic nature of the DCE-MR series. To include the complete series of phases, rather than a single one, we investigate two different approaches. One is to process each phase separately in the CNN and merge the feature maps of the different phases at the end of the CNN. The second approach is to use the phases as different channels of the input image, like an RGB image. Each of the latter two configurations contains identical input information, but it is processed differently, which might result in different outcomes.\par

In this feasibility study, we investigate the performance of two commonly used CNN architectures using the three different input configurations as described above. Both the U-net architecture\cite{Ronneberger2015} and a dilated fully convolutional network (FCN) architecture\cite{Yu2016} were selected, based on their good performance in medical image segmentation tasks\cite{Litjens2017}. The two architectures are compared in order to study the effect of the input configuration regardless of the network architecture.


\section{METHODS}
\subsection{Data}
\label{sec:Data}
The DCE-MR series were acquired in six breath holds with one to five 3D images per breath hold. The DCE-MRI series was acquired on a 1.5 T MRI scanner (Philips, Best, The Netherlands) using a clinical protocol with the following parameters: TE: 2.143 ms; TR: 4.524 ms; flip angle: 10 degrees. After acquiring the first image, gadobutrol (0.1 ml/kg Gadovist of 1.0 mmol/ml at 1 ml/s) was administered at once, followed by 25 ml of a saline solution at 1 ml/s. In total, 16 3D images per patient were acquired with 100 slices and matrix sizes of 256 $\times$ 256. Voxel size was 1.543 mm $\times$ 1.543 mm $\times$ 2 mm. \par
The DCE-MR series of thirty-eight patients from the University Medical Center Utrecht, The Netherlands were used in this work. The data sets were randomly divided in a training set, validation set and test set. Sixteen data sets were used for training, three for validation, and nineteen for testing. The DCE-MR series were corrected for motion using a PCA-based groupwise registration algorithm\cite{Jansen2017}. A zero-mean-unit-variance rescaling was applied to all intensity values between the 0\textsuperscript{th} and 99.8\textsuperscript{th} percentile of the intensity histogram for intensity normalization. In this study we assumed that the 99.8\textsuperscript{th} percentile intensity corresponds to the contrast agent peak in the aorta. Intensities higher than the 99.8\textsuperscript{th} percentile are most probably noise and were assigned the same intensity value as the 99.8\textsuperscript{th} percentile intensity. \par  
The DCE-MR images were combined per breath hold by averaging over the fourth dimension, resulting in six contrast phases of DCE-MRI. The series start with the pre-contrast image. This is followed by the other phases: the early arterial phase, the late arterial phase, the hepatic/portal-venous phase, the late portal-venous/equilibrium phase and the late equilibrium phase.\par
\begin{figure} [t]
	\begin{center}
		\begin{tabular}{c} 
			\includegraphics[width=16cm]{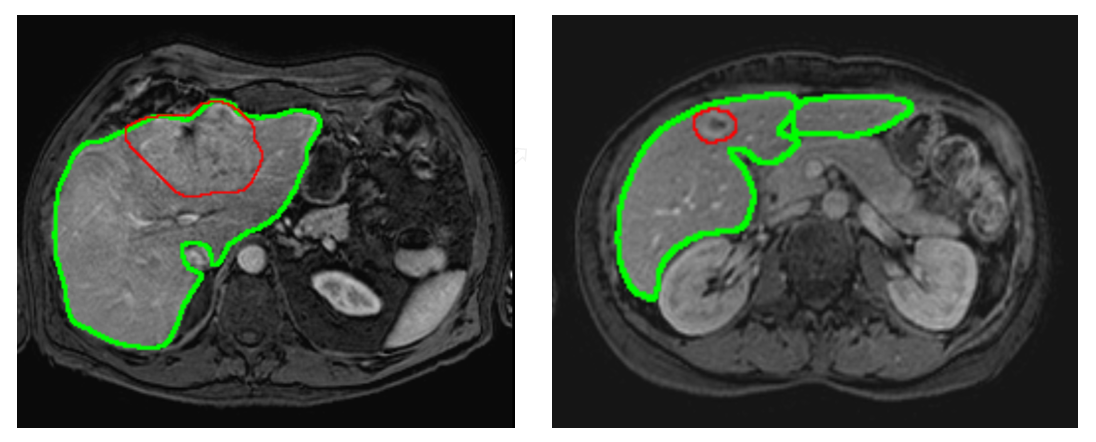}
		\end{tabular}
	\end{center}
	\caption[example]
	{ \label{fig:Examples} 
		 Two examples of the late arterial phase of the DCE-MRI scans. The livers are contoured in green. The lesions are contoured in red.}
		 
\end{figure}

The liver was manually annotated by two observers. The training and test sets were annotated by the first observer. The test set was annotated twice by the first observer with at least one week in between the two annotations and once by a second observer. A radiologist with more than 10 years of experience in liver MR analysis verified all manual segmentations and provided corrections where needed. Most MR data sets had at least one liver lesion, except for three data sets in the training set, one in the validation set and four in the test set. The liver lesions were included in the liver segmentations. Figure \ref{fig:Examples} shows two examples of DCE-MRI scans with a green contour around the liver and a red contour around the liver lesions.

\subsection{Inputs}
Three input possibilities of DCE-MRI for CNNs were evaluated:
\begin{enumerate}[label=\Roman*]
  \item One phase of the DCE-MRI with the best visual contrast between the liver and other abdominal organs (late arterial phase);
  \item All six phases of the DCE-MRI processed by the network one by one and merged at the second last layer of the network;
  \item All six phases of the DCE-MRI as channels of the input image.
\end{enumerate}

\subsection{Network architectures}
\begin{figure} [ht]
	\begin{center}
		\begin{tabular}{c} 
			\includegraphics[width=12cm]{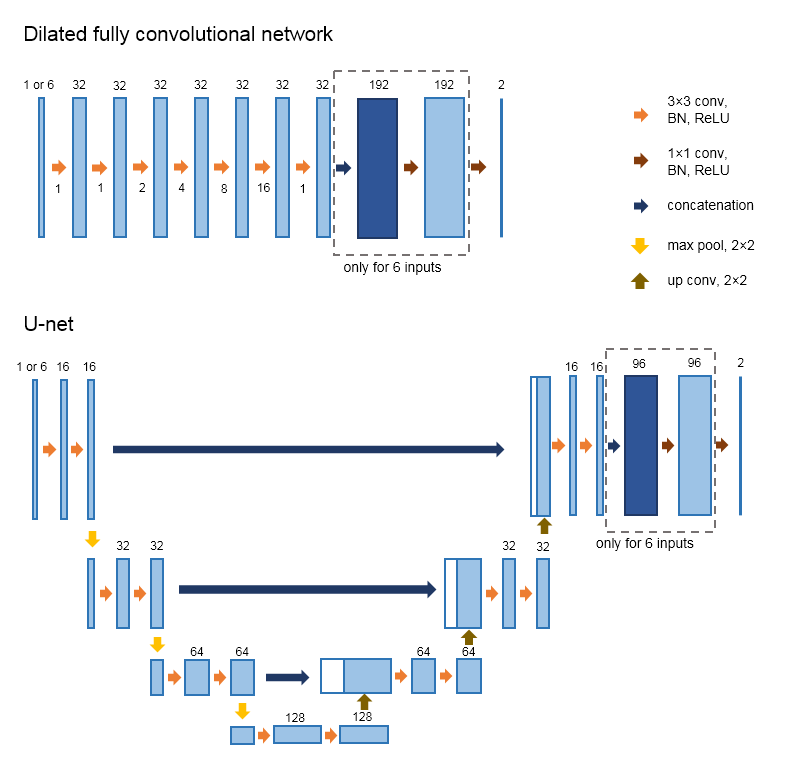}
		\end{tabular}
	\end{center}
	\caption[example]
	{ \label{fig:Architectures} 
		The dilated fully convolutional network and the modified U-net. The dilation rate of the convolution in the dilated FCN is given below the orange arrow, no digit means no dilation. The number of kernels or channels are given above the light blue rectangles. The number above the dark blue rectangle is the size of the merged feature maps. The white boxes in the modified U-net are the copied feature maps.}
\end{figure}

\textbf{Dilated fully convolutional network} The dilated FCN\cite{Yu2016} consists of 7 layers with each a 3$\times$3 convolution and 32 kernels. The 3\textsuperscript{rd} layer has a dilation of 2, the 4\textsuperscript{th} layer a dilation of 4, the 5\textsuperscript{th} layer of 8 and the 6\textsuperscript{th} layer of 16. The final 8\textsuperscript{th} layer has a 1$\times$1 convolution. This gives a receptive field of 67$\times$67 pixels. \\
\\
\textbf{Modified U-net} The second network is smaller than the conventional U-net\cite{Ronneberger2015}, it only has four stages with three skip connections. We choose a smaller network because our images are only 256$\times$256 voxels and an extra layer of downsampling would result in small feature maps. Also the number of kernels was reduced to 16 in the upper layers and to 128 kernels in the bottom layers.\\
\\
For the input approach in which the phase images are used as separate input images (configuration II), the feature maps of the six phases are concatenated and an extra 1$\times$1 convolution layer is added before the final 1$\times$1 convolution layer in both networks, to combine the feature maps. Figure \ref{fig:Architectures} gives an overview of both network architectures.
The loss is calculated by a similarity metric: $(2*X\cap Y+s)/(X^2+Y^2+s)$, where $X$ is the predicted segmentation, $Y$ is the ground truth mask and $s$ is a small number to prevent dividing by zero and set to 1e-5\cite{Milletari2016}. ReLU activation and batch normalization are used in all the convolutional layers, except for the final layer which has a softmax activation. Glorot uniform initialization\cite{Glorot2010} and Adam optimization with a learning rate of 0.001 are used. The networks were trained for 500,000 iterations, with mini batches of one slice per mini batch. No data augmentation was applied.\par
To limit the parameters and training time for this feasibility study, both networks were trained on 2D slices, but the concept can easily be expanded to 3D.

\subsection{Post-processing}
The post-processing consists of setting a threshold of 0.50 on the probability outcome of the networks. After that 3D hole filling is performed, to fill the holes caused by liver lesions. The largest connected component is selected as the final segmentation.

\section{EXPERIMENTS}
The performance of the three input configurations is evaluated based on the Dice similarity coefficient (DSC) and the modified Hausdorff distance (HD) based on the 95$^{th}$ percentile. The first set of liver annotations of the first observer is considered the ground truth. \\
\\
DSC: $(2*X\cap Y)/(X+Y)$, where $X$ is the predicted segmentation and $Y$ the ground truth mask.\\
\\
HD at 95\textsuperscript{th} percentile: $max⁡(h_{95} (X,Y),h_{95} (Y,X)$), with  $h_{95}(X,Y) = K^{95{th}}_{x\in X} min \left \| y-x \right \|$. Where $K$ is the 95$^{th}$ ranked minimum of the minimum Euclidean distances between all points of $y$ and $x$.\\
\\
These two metrics are computed on the segmentation results for each input configuration and network architecture. The DSC and the loss function are both similarity metrics and thus the outcomes of the CNNs are optimized for the DSC. However, the DSC can be used to compare the segmentation results of different network architectures and input configurations, since the loss function is always the same.\\
\\
\textbf{Inter- and intra-observer agreement}\\
The inter- and intra-observer agreements for manual segmentation of the liver in the DCE-MR series in the test set were obtained to compare the performance of the network segmentations with human annotations. The same metrics as mentioned above are calculated for the inter- and intra-observer agreement.\\
\\
\textbf{Statistics}\\
The paired Student's t-test was calculated for the DSC results between the three input configurations, for each of the two networks. Additionally, it was computed between the DSC results of the two network architectures with the six channels image inputs (configuration III). The paired Student's t-test was computed on the DSC results of configuration III and the DSC results of the inter-observer agreement, to test for a significant difference. The same was done for the intra-observer agreement DSC results.\par
The Wilcoxon signed rank test was calculated for the HD results between the three input configurations, for each of the two networks. Additionally, it was calculated between the HD results of the two network architectures for configuration III. To test the difference between the configuration III results and the inter-observer agreement results, the Wilcoxon signed rank test was computed on the HD results. The same was done for the intra-observer agreement HD results.\\

\section{RESULTS}
\begin{figure} [ht]
	\begin{center}
		\begin{tabular}{c} 
			\includegraphics[width=15cm]{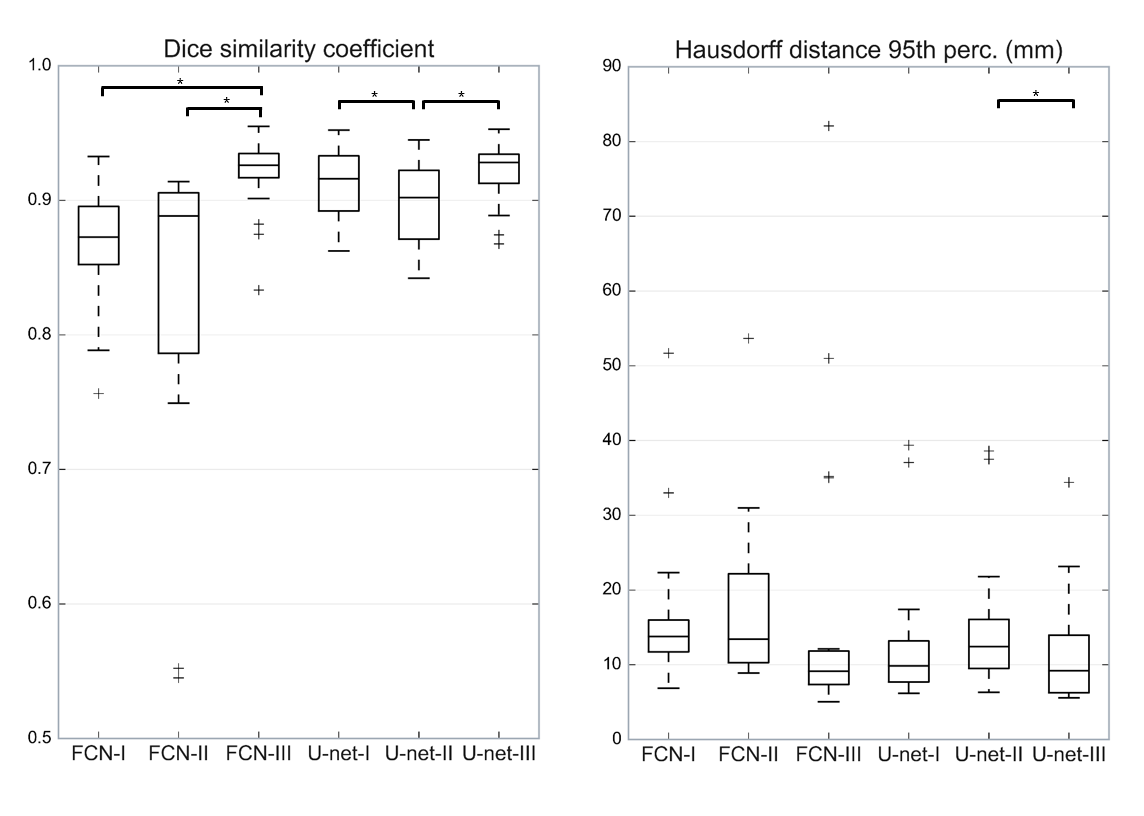}
		\end{tabular}
	\end{center}
	\caption[example]
	{ \label{fig:fig3} 
		Boxplots of the dice similarity coefficient and modified Hausdorff distance (mm) for the dilated FCN and the modified U-net using the three input configurations. The numbers I, II, and III correspond to the three input configurations: one phase image, six phase images, and six phases as channels input image, respectively. An asterisk indicates a significant difference (p$\leq$0.05) between two input configurations.}
\end{figure}

The DSC and HD values of the liver segmentations of the three input configurations for the two CNNs are given in Figure \ref{fig:fig3}. \par
The median [interquartile range (IQR)] inter-observer agreement is 0.943 [0.933 - 0.954] for the DSC and 5.56 mm [4.29 - 6.75 mm] for the HD. The median [interquartile range (IQR)] intra-observer agreement is 0.959 [0.956 - 0.966] for the DSC and 3.09 mm [2.53 - 3.09 mm] for the HD.\par

Paired Student's t-tests show a significant difference (p$\leq$0.05) between the DSC of configuration I vs III for the dilated FCN, between configuration II and III for both the dilated FCN and the U-net, and between configuration I vs II for the U-net.
A significant differences was seen for the HD between configuration II and III for the U-net. No significant difference was seen between the dilated FCN and the U-net architectures when the third configuration is used.\par
The DSC and HD of the computed liver segmentations differ significantly from the DSC and HD of the inter- and intra-observer results in the two metrics for both networks with the best input; configuration III.

\begin{figure} [ht]
	\begin{center}
		\begin{tabular}{c} 
			\includegraphics[width=15cm]{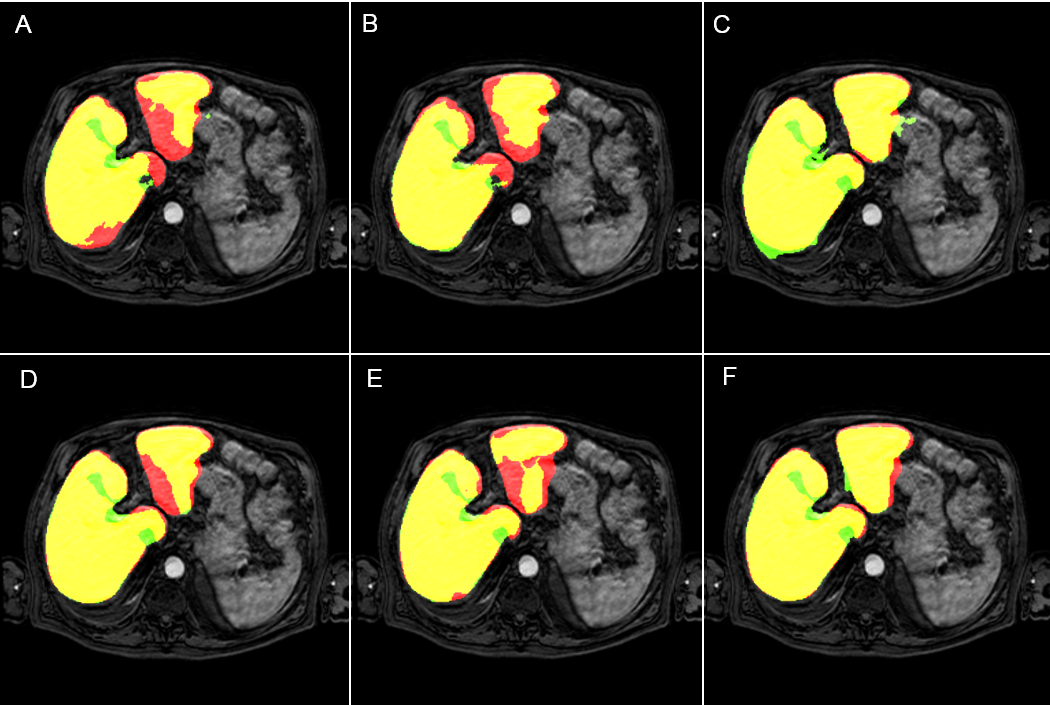}
		\end{tabular}
	\end{center}
	\caption[example]
	{ \label{fig:fig4} 
		Segmentation results after post-processing. The first row represents the dilated FCN and the second row the modified U-net. From left to right the columns represent configurations I, II, and III. Yellow is true positive, red is false negative, and green is false positive.}
\end{figure}

\begin{figure} [ht]
	\begin{center}
		\begin{tabular}{c} 
			\includegraphics[width=15cm]{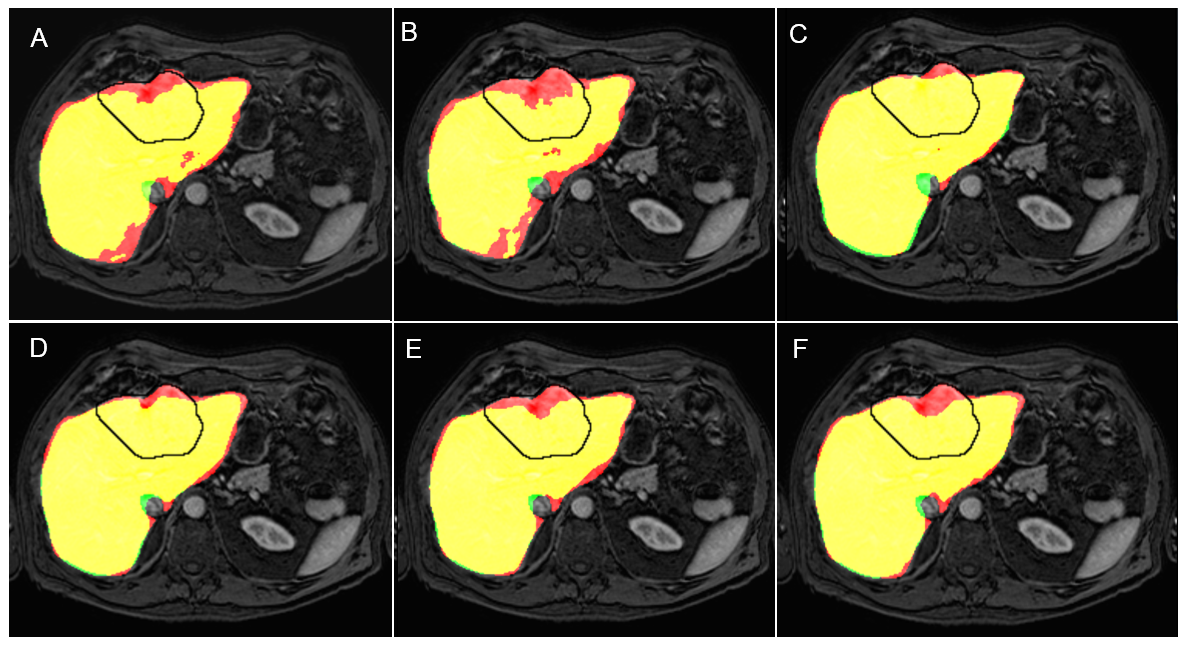}
		\end{tabular}
	\end{center}
	\caption[example]
	{ \label{fig:fig5} 
		Segmentation results of a liver with a lesion, same patient as depicted in Figure \ref{fig:Examples} (left). The first row represents the dilated FCN and the second row the modified U-net. From left to right the columns represent configurations I, II, and III. The liver lesion is contoured in black. Yellow is true positive, red is false negative, and green is false positive.}
\end{figure}

Examples of the segmentation results of one slice for the three input configurations and the two neural networks are given in Figure \ref{fig:fig4}. It shows a typical result, where configuration III shows the best result for both network architectures. Segmentation results of a liver with a large lesion are given in Figure \ref{fig:fig5}. A great part of the lesion is included in the segmentations. Configuration II has the worst results with regard to the lesion for both networks.

\section{DISCUSSION AND CONCLUSIONS}
Combining the different phases of the DCE-MR series in different channels results in significantly better segmentations than combining the feature maps of the different phases at the end of the network. It seems that the network learns the relationship between the phases of the DCE-MR series better when the intensity values still represent the MR intensity, rather than when several convolutional kernels are already applied. Especially the small U-net seems to fail to combine the feature maps at the second last convolutional layer in a proper way. Furthermore, including more information in the input images provides significantly better segmentation results than without the extra information for the dilated FCN.\par 
For both networks, the number of learned parameters is in a similar order of magnitude for the three input configurations, which  means the risk of overfitting is similar for all approaches\cite{Jain2000} and leads to a fairer comparison between the performances of three configurations. However, this meant that the weights of the convolutional kernels needed to be shared in the networks with configuration II. This might not be the most optimal setup, because separately optimizing the kernel weights for each of the six phases of the DCE-MR series might result in better segmentation results. This may explain why configuration II performs less in some cases than configuration I, although more information is provided.\par
The results did not show a significant difference between the two network architectures when using the six phases as channels input image (configuration II). This suggest that the input configuration is more important than the network architecture. Visual inspection showed that, in general, the dilated FCN tends towards oversegmentation, while the modified U-net tends towards undersegmentation of the liver.\par
The extreme outliers of the DSC in the dilated FCN with configuration II are due to severe undersegmentation. The extreme outliers of the modified HD in the dilated FCN with configuration III are due to oversegmentation into either the spleen or the heart.\par
The DSC and HD of the segmentation results differ significantly from the DSC and HD of the inter-observer agreement and intra-observer agreement. Nonetheless, an expert radiologist deemed the segmentations of input configuration III for both networks satisfactory as a rough outline of the liver for further image analysis. Optimizing both networks could boost the performance of the networks. In this feasibility study, both networks provide a rough liver segmentation, which included most of the lesions, except for lesions near the edges of the liver, as can be seen in Figure \ref{fig:fig5}.\par
In conclusion, the configuration of the input images has an important effect on the outcomes of convolutional neural networks. Using the six phases of a DCE-MR series as channels of an input image performs better than using one phase image as input or the six phase images separately as input, regardless of the architecture of the network.

\acknowledgments 
The authors would like to thank FJ Wessels and T van den Beeten for their effort in annotating the data. This work was financially supported by the project BENEFIT (Better Effectiveness aNd Efficiency by measuring and modelling of Interventional Therapy) in the framework of the EU research programme ITEA (Information Technology for European Advancement).

\bibliography{main} 
\bibliographystyle{spiebib} 

\end{document}